# The Accretion Mode in Sub-Eddington Supermassive Black Holes: Getting into the Central Parsecs of Andromeda

C. Alig[1,2,3,4], A. Prieto[3,5,6], M. Blaña[3,4,7], M. Frischman[3], C. Metzl[3], A. Burkert[1,3,4], O. Zier[3,8], and A. Streblyanska[5]
[1] Excellence Cluster ORIGINS, Boltzmannstr. 2, D-85748 Garching, Germany
[2] Leibniz Supercomputing Centre (LRZ), D-85748 Garching, Germany
[3] Universitäts-Sternwarte, Fakultät für Physik, Ludwig-Maximilians-Universität München, Scheinerstraße 1, D-81679 München, Germany
[4] Max-Planck-Institut für extraterrestrische Physik, Gießenbachstraße 1, D-85748 Garching bei München, Germany
[5] Instituto de Astrofísica de Canarias (IAC), E-38200 La Laguna, Tenerife, Spain
[6] Universidad de La Laguna, Dept. Astrofísica, E-38206 La Laguna, Tenerife, Spain
[7] Instituto de Astrofísica, Facultad de Física, Pontificia Universidad Católica de Chile, Av. Vicuña Mackenna 4860, 7820436 Macul, Santiago, Chile
[8] Max-Planck-Institut für Astrophysik, D-85748 Garching bei München, Germany
Received 2023 February 15; revised 2023 June 26; accepted 2023 June 27; published 2023 August 9

## Abstract

The inner kiloparsec regions surrounding sub-Eddington (luminosity less than $10^{-3}$ in Eddington units, $L_{\rm Edd}$) supermassive black holes (BHs) often show a characteristic network of dust filaments that terminate in a nuclear spiral in the central parsecs. Here we study the role and fate of these filaments in one of the least accreting BHs known, M31 ($10^{-7}$ $L_{\rm Edd}$) using hydrodynamical simulations. The evolution of a streamer of gas particles moving under the barred potential of M31 is followed from kiloparsec distance to the central parsecs. After an exploratory study of initial conditions, a compelling fit to the observed dust/ionized gas morphologies and line-of-sight velocities in the inner hundreds of parsecs is produced. After several million years of streamer evolution, during which friction, thermal dissipation, and self-collisions have taken place, the gas settles into a disk tens of parsecs wide. This is fed by numerous filaments that arise from an outer circumnuclear ring and spiral toward the center. The final configuration is tightly constrained by a critical input mass in the streamer of several $10^3$ $M_\odot$ (at an injection rate of $10^{-4}$ $M_\odot$ yr$^{-1}$); values above or below this lead to filament fragmentation or dispersion respectively, which are not observed. The creation of a hot gas atmosphere in the region of $\sim 10^6$ K is key to the development of a nuclear spiral during the simulation. The final inflow rate at 1 pc from the center is $\sim 1.7 \times 10^{-7}$ $M_\odot$ yr$^{-1}$, consistent with the quiescent state of the M31 BH.

*Unified Astronomy Thesaurus concepts:* Interstellar medium (847); Andromeda Galaxy (39); Supermassive black holes (1663); Interstellar filaments (842); Astronomical simulations (1857); Hydrodynamical simulations (767)

## 1. Introduction

The different levels of nuclear activity in galaxies should be related to the availability of circumnuclear material and its ability to lose angular momentum and reach the supermassive black hole (BH). Efficient mechanisms to transport matter to the center include bars (Schwarz 1981; Shlosman et al. 1989) and companion galaxy interactions (Toomre & Toomre 1972). High-angular-resolution observations of nearby active galaxies with the Hubble Space Telescope (HST) have revealed the frequent presence of nuclear dust spirals, lanes, and filaments in the central 100 pc (e.g., Malkan et al. 1998). In some cases, these structures extend over several kiloparsecs across the host galaxy (e.g., Pogge & Martini 2002; Prieto et al. 2019; Leahy et al. 2023 and references therein). The combination of HST optical and adaptive-optics IR images of comparable resolution has revealed the size and morphology of these nuclear filaments in great detail, extending further into the central parsec. Furthermore, their presence is ubiquitous at all levels of nuclear activity and classification. Nuclear dust filaments are seen in highly accreting Eddington sources (ratio of bolometric luminosity to Eddington luminosity, $L_{\rm bol}/L_{\rm Edd} \gtrsim 0.1$), low accreting sources ($L_{\rm bol}/L_{\rm Edd} < 10^{-3}$, Prieto et al. 2014, 2019), and the most inactive (e.g., Elmegreen et al. 1998), including cases such as M31, with $L_{\rm bol}/L_{\rm Edd} \sim 10^{-7}$ (Li et al. 2009; this work) or the Galactic center (Genzel et al. 2010).

This paper investigates the formation, stability, and role of the network of dust/gas filaments surrounding the M31 nucleus. The proximity of M31, 780 kpc, allows us to visualize in great detail the morphology, size, and kinematics of the filaments in ionized gas and dust (Figures 1, 2, and 3). These filaments are seen emanating from a circumnuclear dust ring with a radius between 0.5 and 1 kpc from the center, at an inclination angle of about 45° with respect to the main plane of M31. Block et al. (2006) proposes as the origin of the ring the possible past interaction of M31 with its companion dwarf galaxy M32. Inside the ring, the dust filaments follow circularized orbits around the center, ending in a nuclear spiral in the central hundred parsecs. There is a good correspondence in location and morphology of the most prominent dust filaments with ionized gas (Figure 3) (Jacoby et al. 1985; Ciardullo et al. 1988).

Several papers have attempted to reproduce the morphology and velocity field in the inner kiloparsec of M31, based on either N-body simulations (Block et al. 2006 and references therein) or analytical modeling (Melchior & Combes 2011 and references therein). The reported simulation models provide a snapshot of the filament morphology and velocity field based on specific assumptions, e.g., an interaction of M31 with its companion M32 (Block et al. 2006) or a geometric configuration of the different gas planes in the central region (Ciardullo et al. 1988; Melchior & Combes 2011). In this work, we go a step further by using hydrodynamical simulations to see under which initial conditions the flow of a streamer of gas particles injected from a given radius into the central potential of M31 moves and evolves throughout the inner kiloparsec of the interstellar medium (ISM) of the galaxy to produce both the observed filament morphology and kinematic properties. An extensive exploration of different initial conditions and time intervals in the evolution of the injected streamer is analyzed.

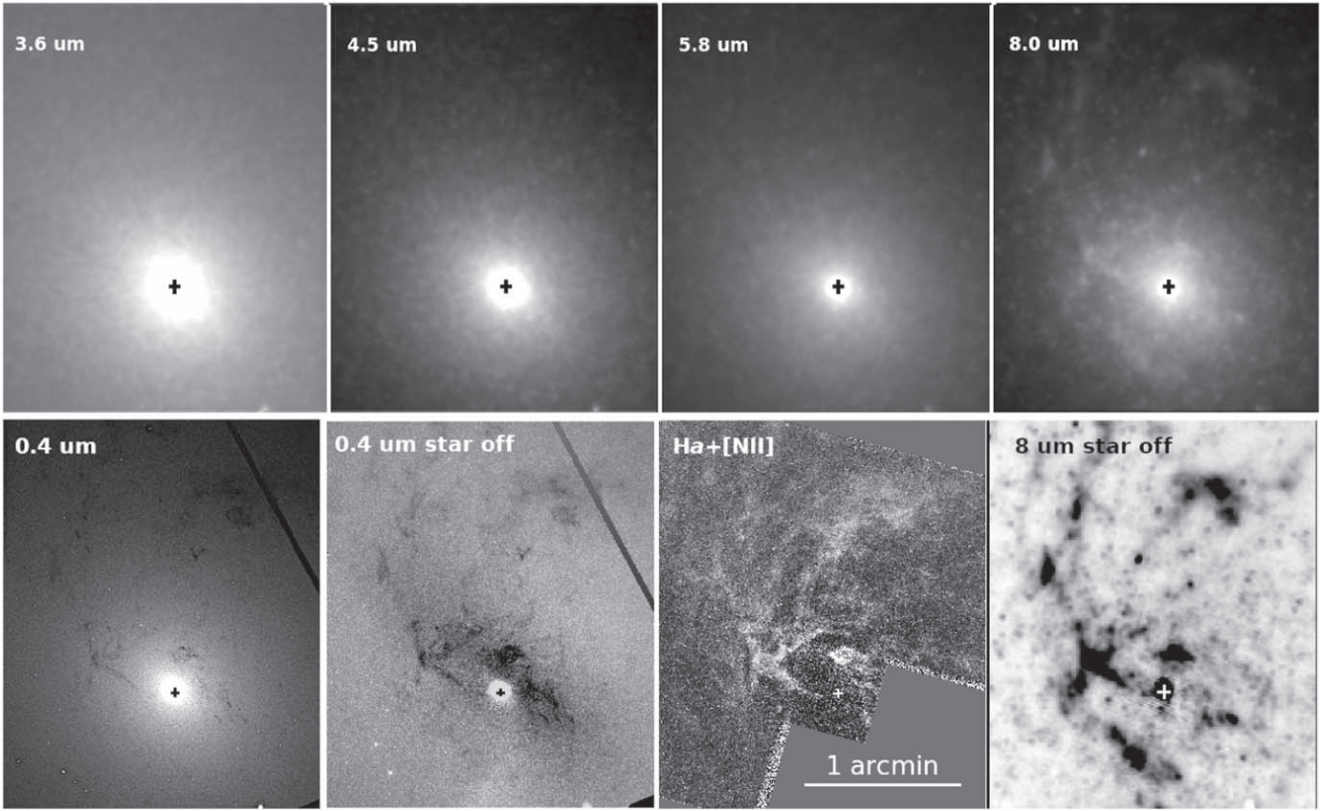

**Figure 1.** Top row: Spitzer images of M31. From left to right, 3.6 $\mu$m, 4.5 $\mu$m, 5.8 $\mu$m, and 8 $\mu$m. The resolution (FWHM) is $1''\!.68$, $1''\!.72$, $1''\!.9$, and $2''\!.1$, respectively. Dust filaments begin to appear in the 8 $\mu$m image. Bottom row: the right panel is the 8 $\mu$m image minus the 4.5 $\mu$m image—the latter taken as representative of the stellar light; the whole filament network is seen in the emission. The middle right panel is the HST-WFPC2-F656N image after continuum subtraction (Section 1.1): the filament network is seen in the ionized H$\alpha$ + [N II] gas emission. The left panel is the UV HST-ACS-F435W image, and the middle left panel is this image after subtraction of the stellar light using a Sérsic profile (Section 1.1); the dust filaments are now seen in absorption as dark structures. The field of view (FoV) of all images is $2' \times 2'\!.5$; the cross is the BH position. North is up, east to the left.

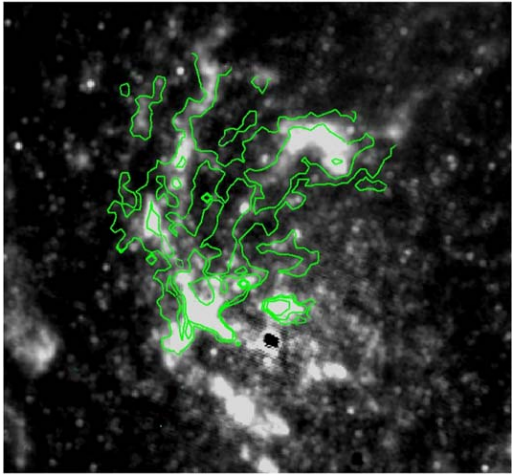

**Figure 2.** The HST H$\alpha$ + [N II] in Figure 1 is shown in contours on top of the 8 $\mu$m star-off image (bottom row rightmost panel) of Figure 1. The FoV is $3'\!.2 \times 3'$. For better visualization, the dust emission is saturated.

By predicting the orbit and velocity of the filaments, we aim to infer the role of the nuclear spiral as a feeder of the M31 BH. Predictions for the mass inflow rate and arrival time of the dust/gas filaments or streamers at the center of M31 are discussed. Various models in the literature that explain the large-scale structure of the M31 galaxy are reviewed in Melchior & Combes (2011) and Blaña Díaz et al. (2018). In this work, a distance to M31 of $0.78 \pm 0.04$ Mpc (de Grijs & Bono 2014) is used, implying $1'' = 3.76$ pc. The Andromeda study forms part of the PARSEC project: "A multiwavelength investigation of the central PARSEC of near galaxies".[9]

### 1.1. Description of the Observations Used in This Work

Mid-infrared Spitzer (Barmby et al. 2006) images of M31 taken with the Infrared Array Camera (IRAC) in 2005 are used in this work. We used corrected Basic Calibrated Data images produced by the Spitzer pipeline, which are corrected for various artifacts such as muxbleed, muxstripe, and pulldown. We further processed, drizzled, and combined these images and associated masks into final mosaics of the central region of M31 (inner four arcminutes) for corresponding IRAC channels centered at 3.6, 4.5, 5.8, and 8 $\mu$m using the standard Spitzer reduction software MOPEX (resolution, FWHM, is $1''\!.68$, $1''\!.72$, $1''\!.9$, and $2''\!.1$ at the four wavelengths respectively). The final mosaics have an integration time (in the circumnuclear region used in our study) of about 45 min.

---
[9] http://research.iac.es/proyecto/parsec/main/index.php

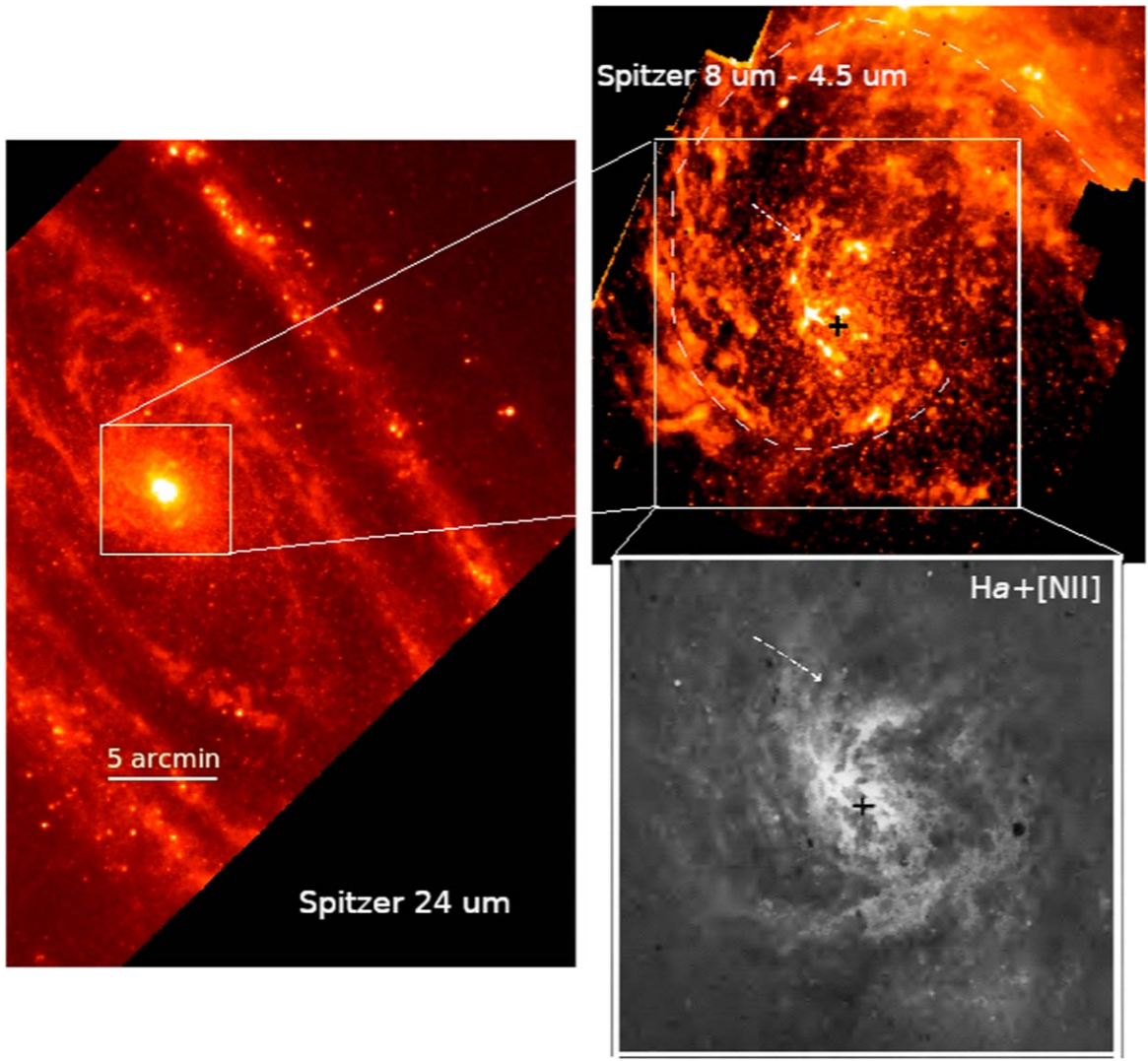

**Figure 3.** Left panel: Spitzer 24 μm image of the central 25′ in the longest direction and 15′ in the perpendicular direction. Top right panel: dust map created by subtracting the Spitzer 4.5 μm image—dominated by stellar light—from the Spitzer 8 μm image—stellar light and dust in emission. The dashed lines in this panel outline the circumnuclear dust ring (see Section 1). The FoV is about 8′.5 × 8′. Bottom right: Hα + [N II] line image from Ciardullo et al. (1988). The white box is 6 × 6 arcmin², approximately the FoV of the Ciardullo et al. (1988) images as derived from cross-matching with the pointlike and dust structures seen in all the images. The arrow indicates the filament used as a reference in the simulations. The BH of M31 is marked with a cross. In all images, north is at the top and east is to the left.

The appearance of the central region of M31 varies dramatically in the different mid-infrared bands, from a smooth featureless bulge dominated by the old stellar population at 3.6 μm to the distinct spiral dust filament structure that dominates the 8 μm image (Figure 1). The dust filaments are also seen with less detail at the longest Spitzer wavelengths (24 μm—Figure 3; 70 μm and 160 μm—shown in Li et al. 2009) due to the degraded spatial resolution in the 6″–38″ (FWHM) range. Thus, in this work we focus on the 3.6 μm to 8 μm images. Those at 3.6 and 4.5 μm are completely dominated by stellar bulge light, while those at 5.8 and 8 μm show a lot of structure caused by dust in emission. A high-contrast image of the dust filaments in the central kiloparsec of M31 is shown in the lower panels of Figure 1 and Figure 3. It shows the 8 μm image after subtracting the close-in-wavelength image that best accounts for the bulge light, for which we use the 4.5 μm image. The result is a pure dust map of the central kiloparsec of M31. It shows a network of narrow dust filaments emanating from a circumnuclear dust ring, following nearly circular orbits around the center. The size of the dust ring as projected on the sky is about 0.5–0.6 kpc radius from the center in the east and south, and about 1 kpc radius in the north. The west side of the ring is not well defined. An equivalent dust map is shown in Melchior & Combes (2011).

The trace of the dust filaments is also visible at shorter wavelengths. The filament pattern is fully revealed in absorption in the HST-ACS UV image (F435W filter) when the stellar contribution is removed (Figure 1). This contribution was determined by fitting a Sérsic model with index 1.35. This was done using the GALFIT code (Peng et al. 2002). It can also be seen in ionized gas, as shown in the bottom panel of Figure 1 with the stellar-subtracted HST-WFPC2-F656N image. The HST-WFPC2-F656N filter is centered on the Hα + [N II] line emission plus stellar continuum light. To remove this light, the adjacent line-free continuum image HST-WFPC2-F547M was rescaled and subtracted from the F656N image.

The spatial coincidence of the ionized gas with the dust filaments is illustrated in Figure 2 with the HST Hα + [N II] image (Figure 1) in contours on top of the dust map (Figure 1, 8 μm star-off). The HST image has a small FoV; only the innermost filaments—a few hundred parsecs—are visible. A much better view of the ionized gas is provided by the Ciardullo et al. (1988) image shown in Figure 3. This image shows in greater detail the entire network of gas filaments in the central kiloparsec. From a morphological point of view, they appear to be circularizing around the center, describing a nuclear spiral.

This nuclear spiral appears to fill the region inside the circumnuclear dust ring (Figure 3), as inferred from the sizes measured directly on the images published in Ciardullo et al. (1988). The ionized gas image with the best contrast by these authors is shown in Figure 3. A close resemblance of the gas filaments to those seen in the dust map is also apparent in Figure 3. It can be seen that the most prominent filaments in the dust map have a counterpart in the ionized gas, albeit with less sharp definition. One obvious similarity is the southern arc that crosses from west to east in the ionized gas image. In the dust map, the arc appears to be a section of the otherwise more extended circumnuclear dust ring. In ionized gas, the arc would be the only visible counterpart to the dust ring.

In comparing the observations with the simulations, we will use the morphology of the dust and ionized gas shown in the images described above. We will also use the velocities reported in the [N II] 6584 Å line emission by Boulesteix et al. (1987), obtained with good spatial coverage by Fabry–Perot observations for the central 400 pc radius. We will also use the molecular gas velocities in CO obtained by Dassa-Terrier et al. (2019) at sparse locations in the central 250 pc radius.

## 2. Setup of the Hydrodynamical Simulations

The following sections describe the parameters and the numerical setup for our hydrodynamical simulations. First, we summarize the general procedure here and go into detail in the specific sections later. To model the central kiloparsec of M31 with our hydrodynamical simulations, we use the M31 potential described in Section 2.2. The simulations cover the range from 900 pc to 6 pc distance from the M31 BH. The starting point is the brightest/longest dust filament, which is seen as a coherent feature extending from about 700 pc north of the nucleus to the central 100 pc (marked with an arrow in Figure 3). The filament curves progressively toward the center as it approaches. It is also seen in the ionized gas, though more diffuse, in the central few hundred parsecs (also marked in Figure 3).

We assume that the filament is formed by gas falling inward from the circumnuclear ring of around 0.7 to 1 kpc radius. We do not investigate how this gas was driven into the ring in the first place—as mentioned in the introduction, one possible mechanism is the interaction of M31 with M32, another possibility is the transport of material along the M31 bar (see Section 2.2). To create the infalling filament, we start the simulation by injecting gas into the potential at a radius of 900 pc. The orientation of the injection plane is taken from a previous analysis by Melchior & Combes (2011), who interpret the observed molecular and ionized gas velocity field as the mixed contribution of an outer gas ring with an inclination of 43° and a position angle of −35°, and an inner disk with a similar inclination but a position angle of 70° (both with respect to the sky plane). For our purposes, we take as the initial configuration for the injection plane of the streamer the outer ring parameters used by these authors. The simulation code and numerical parameters are discussed in detail in Section 2.3 and Section 2.4 respectively. We then let the system evolve for several million years. In the end, we needed about 200 Myr of simulation time to arrive at a configuration that best reproduces the observations, as shown in the results in Section 3.

### 2.1. M31 Galaxy Parameters Used in This Work

The region within the central kiloparsec of M31 is a multiphase medium. According to Li et al. (2009), the absorption-corrected luminosity of the diffuse X-ray gas in the center of M31 as measured by Chandra is $7 \times 10^{37}$ erg s$^{-1}$, and the estimated mass of the hot, $\sim 10^5$ K, gas is $1.2 \times 10^5 \, M_\odot$. The mass in the ionized, $10^{3-4}$ K gas is $10^3 \, M_\odot$ (Rubin et al. 1971). The temperature of the dust in the filaments, inferred from Spitzer photometry in the central 500 pc radius, is in the region of 50 K, and the inferred dust masses are a few thousand solar masses (Li et al. 2009). H I is not detected within 500 pc of M31's nucleus (Brinks & Shane 1984), but several CO clumps are detected in the central 250 pc, all adding up to a few $10^4 \, M_\odot$ (Dassa-Terrier et al. 2019) in molecular gas. A few CO clumps are also detected in the circumnuclear ring (Melchior & Combes 2011). The width of the dust filaments is in the range of 1 to 3 pc, as measured in the UV HST-ACS-F435W image, where the filaments are seen in absorption against the background starlight (Figure 1, also in AstroSat UV images in Leahy et al. 2023).

The accretion luminosity of M31 is derived by integrating the subarcsecond-resolution spectral energy distribution (SED) of the region containing the P2 + P3 peak emissions surrounding the BH (Bender et al. 2005). We proceed with aperture photometry in HST images in a 0″.25 radius surrounding both peaks (Alvarez, Prieto & Streblyanska in preparation). The SED spans the range 0.3–1.6 μm and the integrated luminosity from this SED is $\sim 2 \times 10^{39}$ erg s$^{-1}$. Assuming that this luminosity represents an upper limit to the BH accretion luminosity ($L/L_{\rm Edd} < 10^{-7}$), it is one of the lowest along with that of Sgr A*.

### 2.2. M31 Potential

We use as input for the gravitational potential the M31 dynamical models of Blaña Díaz et al. (2018). These authors use the made-to-measure method to construct self-consistent equilibrium dynamical models that are fitted to the 3.6 μm Spitzer images, the stellar kinematic observations from the VIRUS-W IFU observations (Opitsch et al. 2018; Saglia et al. 2018; Gajda et al. 2021), and the H I rotation curve of the disk (Corbelli et al. 2010). The modeling explores dark matter halos with Navarro–Frenk–White and Einasto profiles. Here we use their best-fit model JR804, which uses an Einasto profile. They determined masses for each component of the M31 composite bulge, finding a classical bulge dominating mass in the center (see also Blaña Díaz et al. 2017), and a bar with a box/peanut bulge substructure dominating stellar mass in the outer region of the bulge. From this analysis, a bar pattern velocity frequency of 40 km s$^{-1}$ kpc$^{-1}$, or equivalently a rotation period of 154 Myr, is estimated. We input the gravitational potential of model JR804 into the hydrodynamical simulations as a 3D Cartesian grid of acceleration with a resolution of 10 pc within a cube of 2 kpc side length.

In addition to the stellar and dark matter gravitational potential of M31, we add the point-mass potential of the $1.4 \times 10^8 \, M_\odot$ BH and the nuclear stellar disk surrounding the BH, which has a mass of about $1.0 \times 10^7 \, M_\odot$ (Tremaine 1995; Bender et al. 2005). The radius of the nuclear disk is about 6 pc, which also corresponds to our inner simulation boundary. Particles falling below this boundary are removed from the simulation and counted as accreted. Of course, accretion here does not mean direct accretion onto the BH, but only material falling into the central 6 pc. The low spatial resolution of the simulations for the innermost BH region and the small particle time steps required to simulate gas close to the BH led us to exclude this region in order to speed up the simulations and focus on the more numerically robust results for the outer region. We note that values for the mass of the M31 BH can range from a few $10^7$ to a few $10^8 \, M_\odot$ (Al-Baidhany et al. 2020), but test simulations have shown that the influence of the BH on orbits outside the 6 pc accretion radius is negligible below a few $10^9 \, M_\odot$, so the value of the BH mass we choose is not very important.

### 2.3. Setup of the Simulations: Code Implementation

We perform our hydrodynamical simulations using the *N*-body smoothed particle hydrodynamics (SPH) code GADGET3 (Springel 2005), which uses the entropy-conserving formulation of SPH. The code incorporates a hybrid OpenMP/MPI approach, which is well suited for simulations where the balances of memory and workload can easily diverge, which happens in our simulations due to the continuous injection of gas into the simulation domain. The code base is taken from the repository of Volker Springel, so except for a few modifications described below, we use the mainline version of GADGET3.

In recent years it has been found that the standard implementation of SPH has problems in handling contact discontinuities and mixing of fluids in different phases (Agertz et al. 2007; Wadsley et al. 2008). As a result, traditional SPH is unable to fully reproduce fluid instabilities, so we take advantage of a number of SPH improvements implemented in GADGET3 by Beck et al. (2016).

The first is a higher-order kernel. The standard implementation uses the cubic spline (Monaghan & Lattanzio 1985), which requires about 64 neighbors to estimate particle density. The improved implementation uses the Wendland kernel family, which achieves better numerical convergence. In our case, we use the Wendland $C^4$ kernel, which requires about 200 neighbors to estimate the particle density. Of course, due to the larger number of neighbors required, the computational cost increases.

The next improvement replaces the original artificial viscosity description in GADGET3. SPH solves the Euler equations, which neglect dissipative terms such as viscosity and thermal dissipation. To include viscous effects in the the standard SPH implementation, an additional term is added that converts kinetic energy to thermal energy, with the strength of the conversion modeled by a given constant (Monaghan 1997). However, because a single constant determines the strength of viscosity throughout the simulation, viscosity is applied with equal strength in places where strong viscosity is desired (strong shocks) as well as in places where low viscosity is desired (rotating and shearing flows). A model that recognizes the need for adaptive viscosity strength has been developed by Cullen & Dehnen (2010). In this model, shocks are identified and the strength parameter for the artificial viscosity is tuned for each individual particle. Similarly, an advanced implementation for thermal conduction is added using the model of Tricco & Price (2013), which, similar to the artificial viscosity model, can tune the conduction strength for individual particles.

The standard SPH implementation already recognizes the need to reduce the artificial viscosity in shear flows or rotating systems, because systems such as gas disks can quickly break apart if the full-strength artificial viscosity is applied. However, the shear viscosity limiter often used in the standard SPH implementation (Balsara 1995) uses only lower-order terms in estimating the curl of the velocity field to get a handle on the strength of rotation. The implementation of the shear viscosity limiter of Price (2012) uses higher-order terms and is therefore also stable in strong shear flows.

Finally, we added a module that allows the creation of new gas particles during the simulation. To create a particle, a random position within the target region is generated and the nearest neighboring particle is determined. The new particle is then created as a clone of the existing particle and the basic properties (position, velocity, temperature) of the new particle are updated. If the new particle position is closer to the nearest neighbor than about three times the minimum SPH smoothing length, it is discarded and a new position is created to ensure that we do not go below the regime in which we can resolve the particle–particle forces. The new particle is always placed in the lowest available time bin so that all SPH properties are updated as quickly as possible. To ensure equal SPH particle masses, particle injection occurs only when the total accumulated mass to be injected at a given rate is greater than the mass of a single SPH particle. If the rate is high enough, it is also possible to inject multiple particles in a single time step.

### 2.4. Initial Conditions and Numerical Parameters

Particles are injected with an initial temperature of $10^4$ K and cooling is performed via the cooling curve adapted from Smith et al. (2017) assuming solar metallicity. The minimum allowed temperature is set to 50 K, which is approximately the inferred temperature of the dust in the filaments (see Section 2.1), but no particle ever reaches this floor during the entire simulation. For the gaseous component we use an adiabatic equation of state with an adiabatic exponent of $\gamma = 5/3$ and a mean molecular weight of $\mu = 1.27$, corresponding to a standard mixture of 71.1% atomic hydrogen, 27.41% helium and 1.49% metals.

The number of SPH neighbors is set to 200 to account for the higher-order (Wendland $C^4$) kernel. Our minimum value for the adaptive SPH smoothing length as well as the fixed gravitational softening length is set to 0.01 pc. The Courant number is set to 0.2 for all simulations. The injection region is cylindrical with a radius of 2.5 pc and a length of 5 pc (corresponding to the filament width seen in the absorption, Section 2.1), but due to thermal expansion the stream quickly expands to a radius of about 30 pc, which is comparable to the width of the filament bundles seen in the emission in the Spitzer images. Using a small injection region to allow for some thermal expansion compensates for the effect of the random selection of the injection position and naturally replaces the random distribution with a more glass-like configuration created by the repelling forces. The injection velocity is set to $-100$ km s$^{-1}$ and the injection angle is $-40°$ against the *x*-axis of the injection plane. The initial position of the center of the injection cylinder is set to 900 pc on the *y*-axis of the injection plane.

To produce a filament with a length comparable to observations, the injection is stopped after 40 Myr, which was determined by several test simulations. An injection rate of $10^{-4}$ $M_\odot$ yr$^{-1}$ is used as a fiducial value, leading to a maximum total mass (ignoring accretion) of 4000 $M_\odot$ within the simulation domain. These parameters are chosen as the best compromise after running several test simulations. We find that injection of lower total masses ($<10^3$ $M_\odot$) leads to diffuse filaments that quickly disperse, while higher total masses ($>10^5$ $M_\odot$) lead to fragmentation with subsequent star formation, which is not observed. In addition, the injection velocity was also tuned using test simulations. The injection velocity is constrained to be less than the escape velocity, but large enough so that the initial particle orbits do not fall radially into the 6 pc accretion region. The chosen parameters are reasonable and not too extreme for the galactic center of M31. Finally, the mass of a single SPH particle is set to $10^{-3}$ $M_\odot$ so that the maximum number of particles is $4 \times 10^6$.

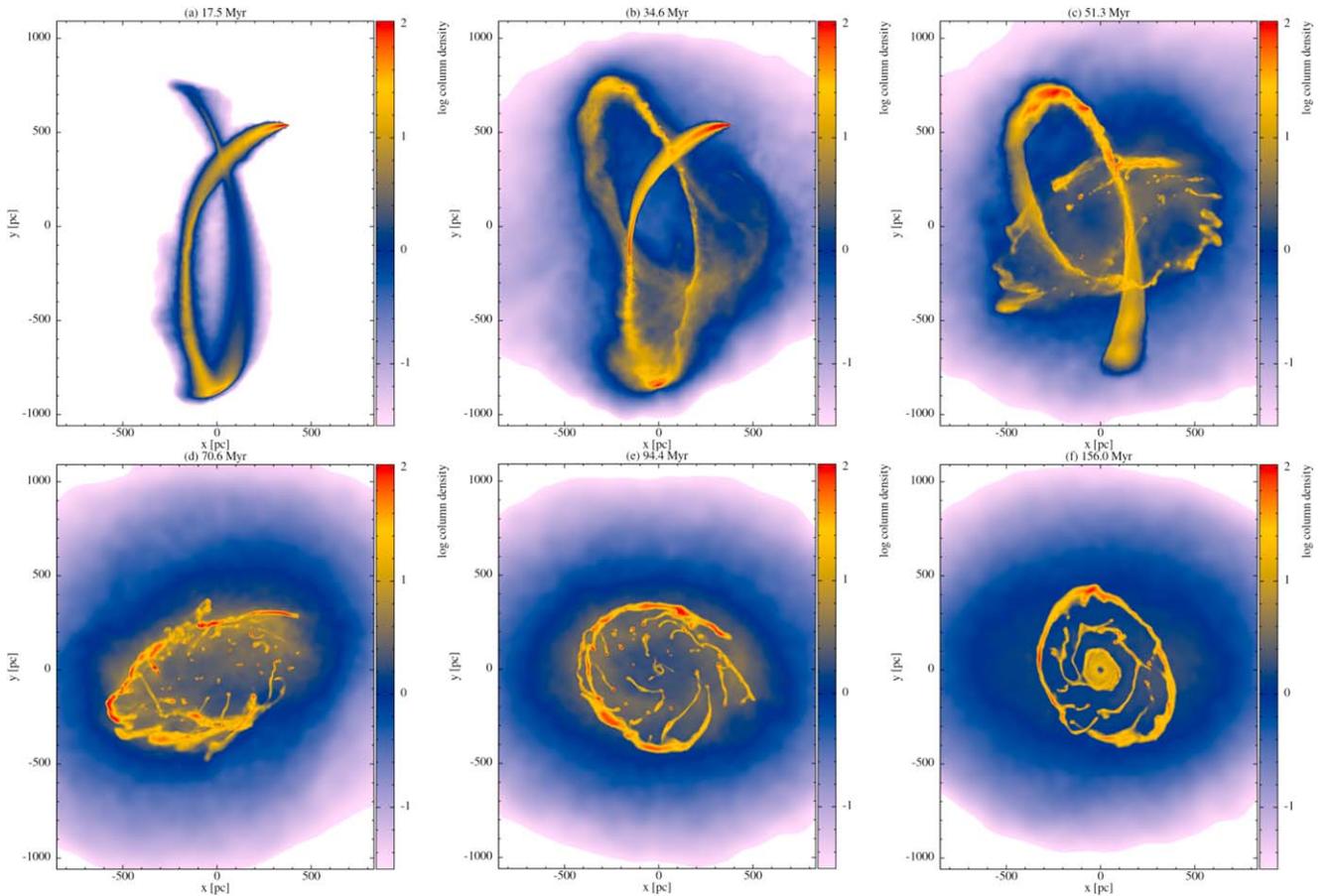

**Figure 4.** Snapshots of the simulation surface density in units of $M_\odot$ pc$^{-2}$ in chronological order. The first few windings of the stream, in panels (a) and (b), do not show much deviation from a pure *N*-body orbit. However, in (b) material from the uppermost arc collides with the freshly injected part of the stream, heating the gas to several $10^6$ K. This leads to the formation of a hot atmosphere, which is seen as the diffuse background from (b) to (f). As the stream continues to cross itself and experience friction within the atmosphere, it slowly circularizes, as seen in the sequence from (d) to (f). Friction at the inner edge of an elongated ring structure that forms in (e) causes thin filaments to spiral inward, eventually forming a small disk in the inner 100 pc, visible in (f).

## 3. Results

### 3.1. Simulation Evolution

We follow the evolution of the surface density (in units of $M_\odot$ pc$^{-2}$) of the stream in Figure 4. The injection point is the upper right nozzle with a higher surface density due to the small size of the injection cylinder (Section 2.4). Initially, the stream stays close to the orbit that would be expected for a pure *N*-body simulation. In panel (a), when the stream crosses itself, the leading part actually passes under the freshly injected part of the stream. Deviations from the pure *N*-body orbit begin to appear when the stream, after forming the upper left arc in (b), falls back toward the freshly injected part. The stream has now broadened considerably and partly collides with its two still well-defined younger loops. This interaction strongly heats up the interacting parts, causing gas of several $10^6$ K to spread out into the surroundings, forming an atmosphere surrounding the region. In panel (c) the injection has already been turned off (at ~40 Myr, Section 2.4), and the trailing part of the stream is now tracing the initial orbit one last time. The leading part visibly breaks up in (c) and starts to circularize in (d) to (f), due to the continuous self-crossing of the stream and the onset of friction with the hot atmosphere. In (e) the friction with the hot atmosphere at the inner edge of the ring leads to instabilities that can break off and spiral into the center, eventually forming an inner disk of about 100 pc radius as seen in (f).

We used the full temperature range of the gas in the simulation to create Figure 4. The injected stream consists mainly of warm gas at a few $10^4$ K, with dense parts that can cool below $10^4$ K and that will be used for comparison with CO observations in the next section. The interstellar medium consists mainly of $10^6$ K gas produced by the self-intersection of the stream. After the injection stops at 40 Myr and the total mass changes only due to accretion, the temperature (*T*) distribution at 51.3 Myr (Figure 4(c)) is a mix of 11.72% in gas at $T < 10^4$ K, 36.35% in gas at $10^4$ K $< T < 10^5$ K, and 51.93% at $T > 10^5$ K. This distribution remains almost constant until 94.4 Myr (Figure 4(d)), when it reaches 12.23%, 41.79%, and 45.98% for the three temperature ranges. In the last panel (Figure 4(f)) the distribution settles at 6.7%, 39.0%, and 54.3%.

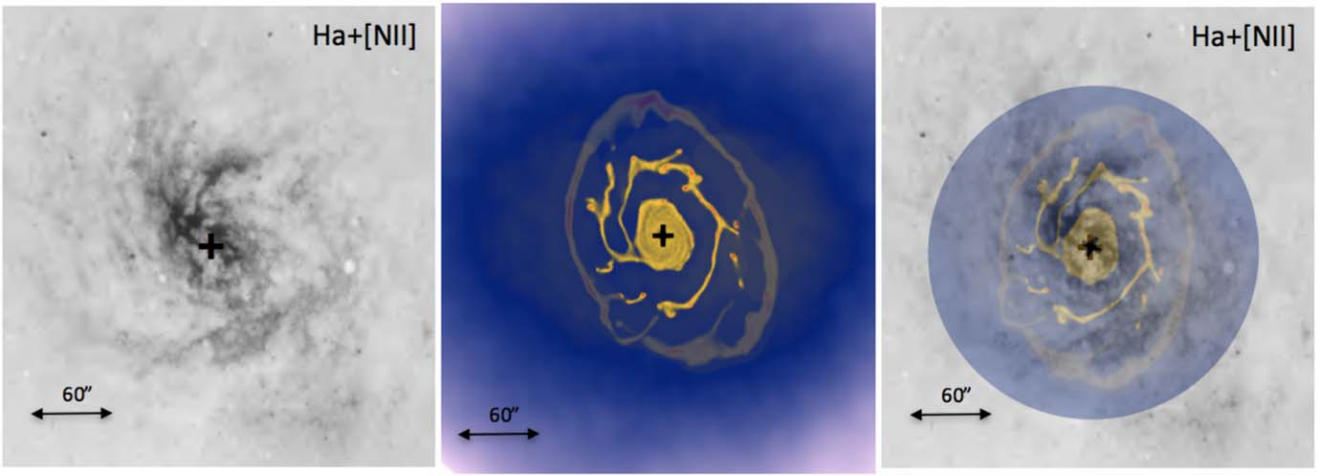

**Figure 5.** Central 1.35 × 1.35 kpc² region of M31; the BH is located at the cross in the center of the panels. The left panel is the Hα + [N II] image adapted from Ciardullo et al. (1988). The central panel is the simulated surface density after 160 Myr of evolution from the injection of the particle streamer. The panel is extracted from Figure 4(f) and corresponds to the same physical region as the left panel. The outer ring is blurred into the background for better comparison of the inner region with observations. The right panel shows an overlay of simulation and observation, with the simulation image slightly rotated to better match the orientation of the observational image.

Comparing the state of the simulation reached at 156 Myr, Figure 4(f), with the morphology (and later in Section 3.2 the velocity field) of the ionized gas in Figure 5, we find particular agreement between observation and simulation shown by the dense disk formed in the inner roughly 150 pc region and the existence of multiple filaments that can be seen feeding the central disk while spiraling inwards. The spiral resulting from simulations is somewhat smaller than the observed one, so the arc feeding into the disk is closer to the inner region than in observations. Still, the overall shape reproduces observations well.

The outer ring seen in the simulation (which is blurred into the background in Figure 5) at 400–500 pc radius has no visible counterpart in the observations at the scale at which we find it. However, taking a close look at both Figure 5 and the dust map in Figure 3 shows that the southern arc in observations actually connects back to the 1 kpc circumnuclear dust ring of Andromeda. In the simulation the southern arc also connects back to the 400–500 pc ring, thus our configuration reproduces observations but on a smaller scale, with the 400–500 pc ring corresponding to the 1 kpc circumnuclear ring of M31. We speculate that changing our initial conditions by starting the injection further out would also lead to an increased size of our ring and subsequently the inner spiral, thus reproducing the correct scale of the observed M31 configuration. The simulations were run up to 200 Myr to verify that the configuration found at 156 My was stable. Over this elapsed period of 200 Myr, only about 33 $M_\odot$ were accreted into the inner simulation boundary of 6 pc. This corresponds to an inflow rate, averaged over the entire period, of $\sim 1.7 \times 10^{-7}\ M_\odot\ {\rm yr}^{-1}$, which agrees with the quiescent state of the M31 BH. Interestingly, this inflow rate is of the same order as that inferred from the ratio of the M31 luminosity to the Eddington luminosity, $L/L_{\rm Edd}$ (see Section 2.1).

The formation of the hot, $10^6$ K, atmosphere plays an important role in the evolution of the system. To show the heating effect due to self-intersection of the injected stream as it moves in the ISM, we follow the motion of a small patch of gas, 16 pc in length, marked as a green square in Figure 6. The patch moves on the $x$-axis from a location at about −220 pc toward the dense, freshly injected part of the stream at about −180 pc in the simulation box.

In Figure 7 the particles within the patch are plotted color-coded for different epochs of the evolution (black then red then blue in steps of 0.1 Myr). The $x$-axis corresponds to the $x$-position of the particle, the $y$-axis shows the temperature. At the last time step (blue), a fraction of the gas has heated up significantly. Also visible is the strong backsplash of the heated gas in the direction opposite to the flow.

To demonstrate the influence of the hot atmosphere on the global evolution, we ran a simulation in which all gas above a temperature of $10^6$ K is removed from the simulation during runtime as soon as it is formed. The corresponding evolution of the surface density is shown in Figure 8 at exactly the same time epochs as shown in Figure 4. Overall, the evolution now resembles a pure N-body simulation (overplotted in white in the first three panels). In the final state, the filamentary structures flowing into the central 100 pc from all sides in Figures 4(e) and (f) have now disappeared.

### 3.2. Line-of-sight Velocity

Another test of the simulations is to see whether the derived velocities agree well with the observed gas velocities. To do this, we extracted the CO line-of-sight velocities for the inner region of M31 from Dassa-Terrier et al. (2019). These velocities are shown as color-coded blobs in the left panel of Figure 9, representing the central region of ∼200 pc radius measured by these authors. An additional CO velocity measurement was extracted from Melchior & Combes (2011) (shown in the right panel of Figure 9) at a location about 300 pc north of the M31 center (marked with a dashed arrow in both panels of Figure 9), which the authors identify as M31A. This region coincides with a prominent dust feature, a concave arch in the dust map.

Unlike Dassa-Terrier et al. (2019), Melchior & Combes (2011) did not detect CO at radii smaller than 300 pc.

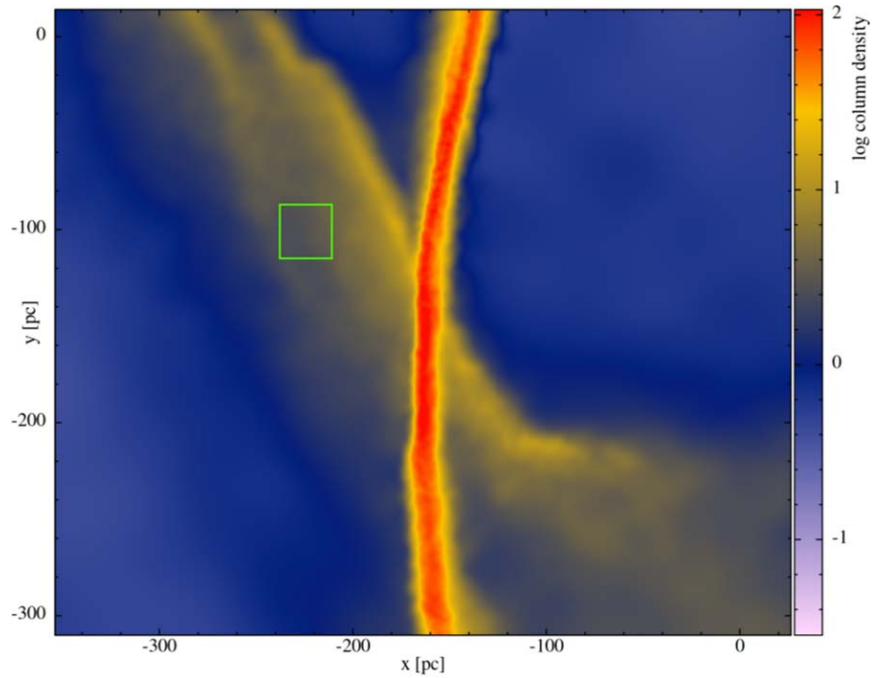

**Figure 6.** Zoom-in on the simulation surface density in units of $M_\odot$ pc$^{-2}$ at 34.6 Myr. Marked in green is the patch of gas that we follow in Figure 7 to demonstrate the heating effect due to the self-intersection of the stream.

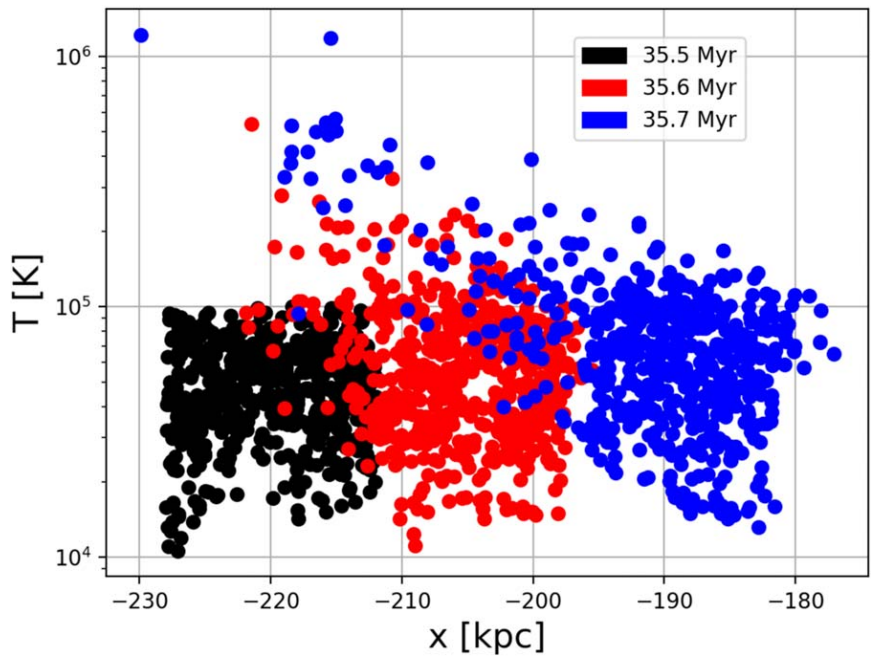

**Figure 7.** Temperature evolution of the gas patch marked in Figure 6. The x-axis shows the particle x-position and the y-axis the temperature. The SPH particles of the first snapshot at 35.5 Myr are plotted in black, the second and third snapshots in red (35.6 Myr) and blue (35.7 Myr). The strong temperature increase in the last time step also leads to a backsplash of the particles in the direction opposite to the flow.

Therefore, we decided to limit our comparison to the set of measurements by Dassa-Terrier et al. (2019), which provide a fair spatial sampling of the central 200 pc, and to complement this set with the measurement of the 300 pc blob because of its CO detection with high signal-to-noise ratio. CO measurements at further radii by Melchior & Combes (2011) are not included in the comparative analysis, as their kinematics show complex behavior and often differ from the velocities of ionized gas measured at comparable locations. The inner 100–200 pc region on the right panel of Figure 5 displays a change in the rotation pattern. The full set of velocities used for the comparison is shown in Figure 9, including the systemic velocity of M31 of about $-300$ km s$^{-1}$ (Braun & Thilker 2004).

The comparison with simulations is shown in Figure 10 at different times in the evolution. The observed CO velocities are shown on top. Starting from the upper left panel of Figure 10 and following the time evolution, one immediately sees that the plane in which the simulated stream evolves is precessing. Thus, observation and simulation are expected to agree only over a limited period of time (if at all). For example, at 135.2 Myr (lower left panel) the simulated central velocities are in good agreement with the data, but not the region 300 pc north. Conversely, at 170.5 Myr, the velocity in this northern region is reproduced, but the central velocity field is not. In between, the evolutionary state reached at ∼160 Myr seems to have features that partially match both the central and northern velocity measurements. Thus, it becomes clear that the northern region velocity of about −150 km s$^{-1}$ cannot be fitted simultaneously with the central velocity pattern.

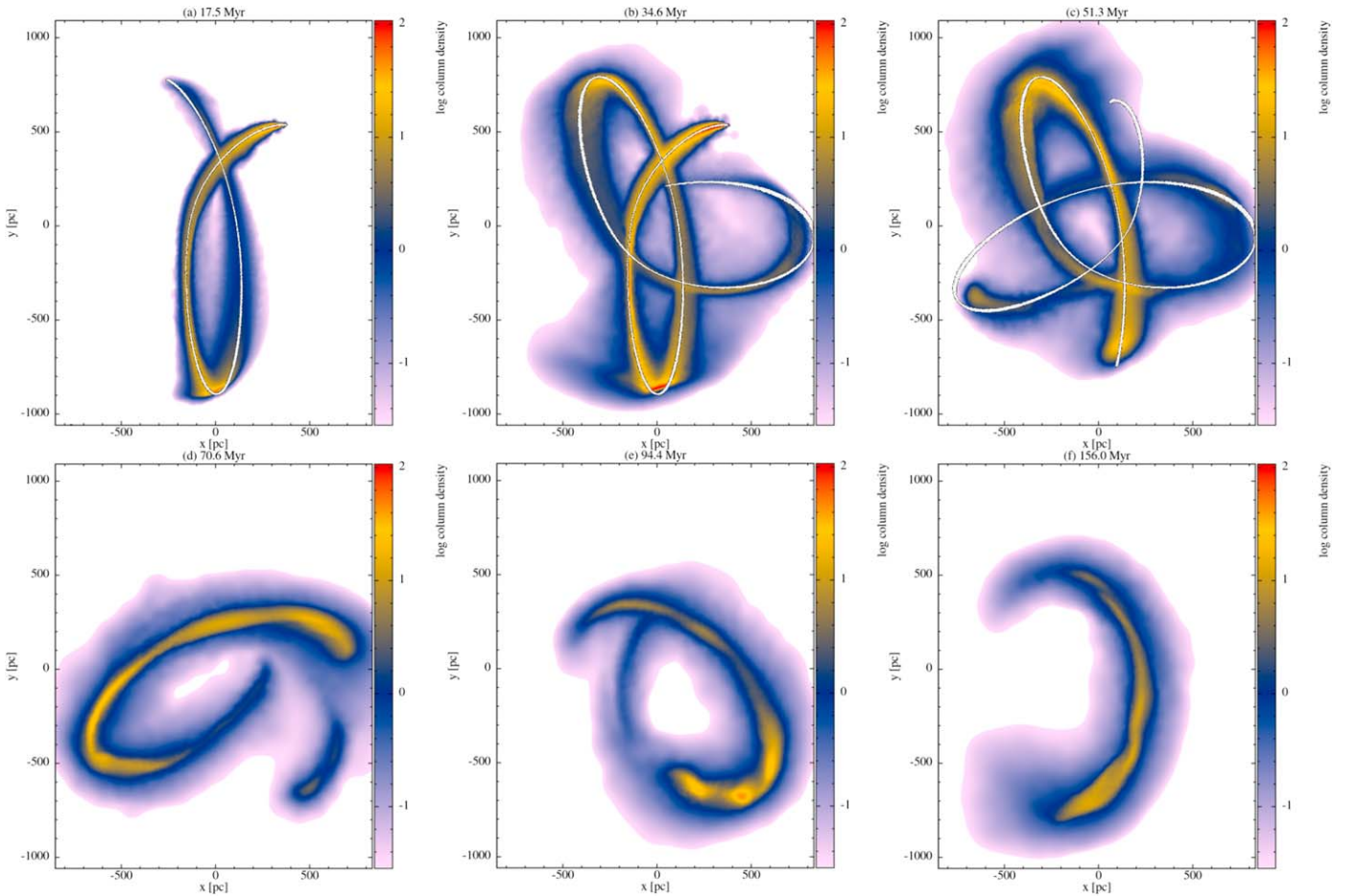

**Figure 8.** Snapshots of simulation surface density in units of $M_\odot$ pc$^{-2}$ in chronological order, with all gas above $10^6$ K removed during runtime. The snapshots are taken at the same times as in Figure 4. Initially, the simulation evolves similarly to the case of a hot atmosphere in Figures 4 in (a) and (b), but it already starts to deviate in (c) due to the absence of the background atmosphere. The system now follows more closely the expected orbit of a pure *N*-body simulation at all times. For comparison, we have plotted the *N*-body orbit in white for the first three panels. At intersections of the stream with itself, parts of the stream are completely removed, leading to the large gaps seen in (d). In the end, only a small part of the filament is left, which slowly circularizes.

On the basis of the asymmetric morphology of the H$\alpha$ + [N II] nuclear spiral in the center of M31, Ciardullo et al. (1988) suggested that the gas in the central region may be in a warped disk, with the southern region seen closer to face-on than the northern side. We noted in Section 1.1 that the dust ring, and hence the distribution of ionized gas within it, is asymmetric, with the south and east radii being a factor of two smaller than the north radius. We also noted that the velocity field of the ionized gas in the central 200 pc or so is decoupled (possibly due to a different rotation angle or disk inclination) from that of the surrounding gas (right panel of Figure 9). However, in our simulations the gas in the central kiloparsec is distributed in a single plane. Thus, the possibility of a different geometrical distribution for the gas in the central few hundred parsecs, e.g., in different planes or a warped region, and the physical causes that could lead to it, will be explored in the next section.

### 3.3. Origin of an Inner Tilt Disk

In the previous sections we have shown the simulation results for a large parameter study including different initial angles, velocities, distances, and user-defined orientations of the injection plane. Interestingly, due to the relatively good radial symmetry of the M31 potential in the inner 1 kpc, all simulations lead to very similar results, equivalent to those shown in Figure 4. Of course, the plane of the resulting disk may differ, as well as the size and asymmetry, and details in the morphology. Nevertheless, the overall results capture most of the important features in the observations and are robust to changes in the initial simulation parameters, as long as they are kept reasonable with the expected conditions in the center of M31. However, it is clear that our basic model alone cannot account for a possible tilt or warp between the inner and outer regions of M31, since the simulations so far always end up with a single precessing plane.

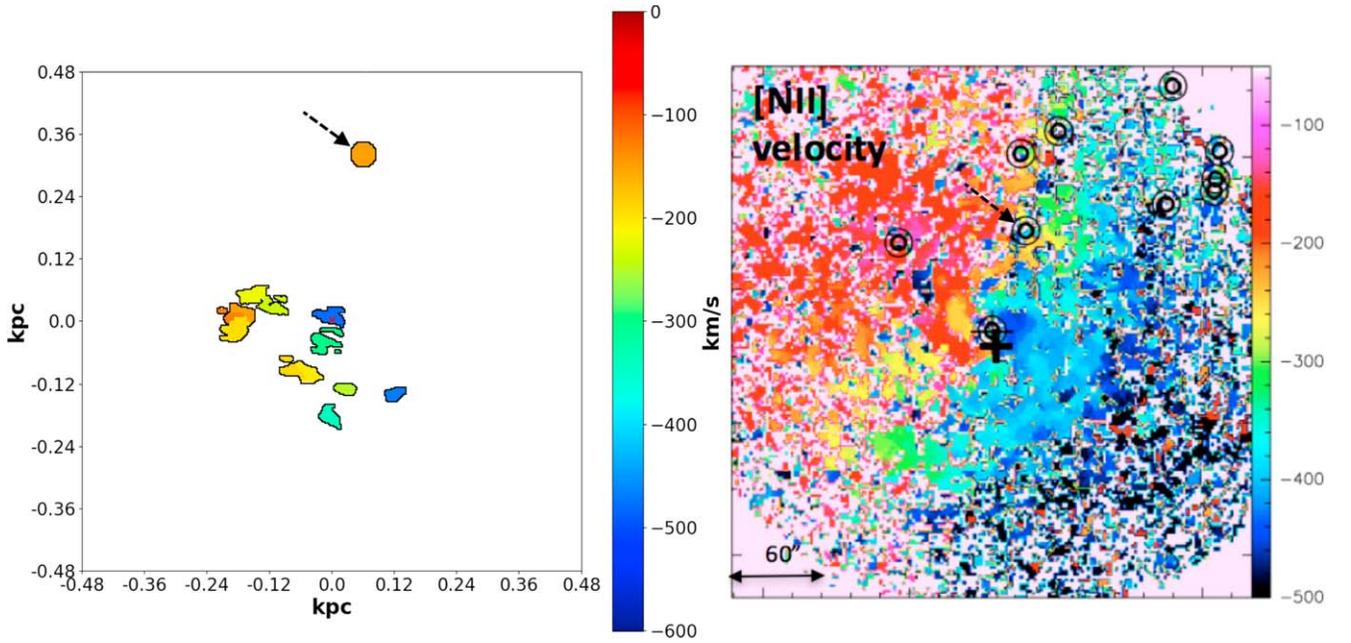

**Figure 9.** Left panel: available CO line-of-sight velocity data within the central 0.5 kpc radius of M31. Velocities include the M31 systemic velocity of about $-300$ km s$^{-1}$. The data are from Dassa-Terrier et al. (2019) (central 200 pc) and are complemented by an additional velocity measurement by Melchior & Combes (2011) about 300 pc north, their region M31A. The BH is at the coordinate origin (red ×). Right panel: the velocity field extracted by Boulesteix et al. (1987) from a Fabry–Perot [N II] 6584 Å image. The image is taken from Melchior & Combes (2011); the circles at the top of the image mark the beam size of the CO observations by these authors and the black hole is marked as a black cross. The M31A region is marked with a dashed arrow in both panels.

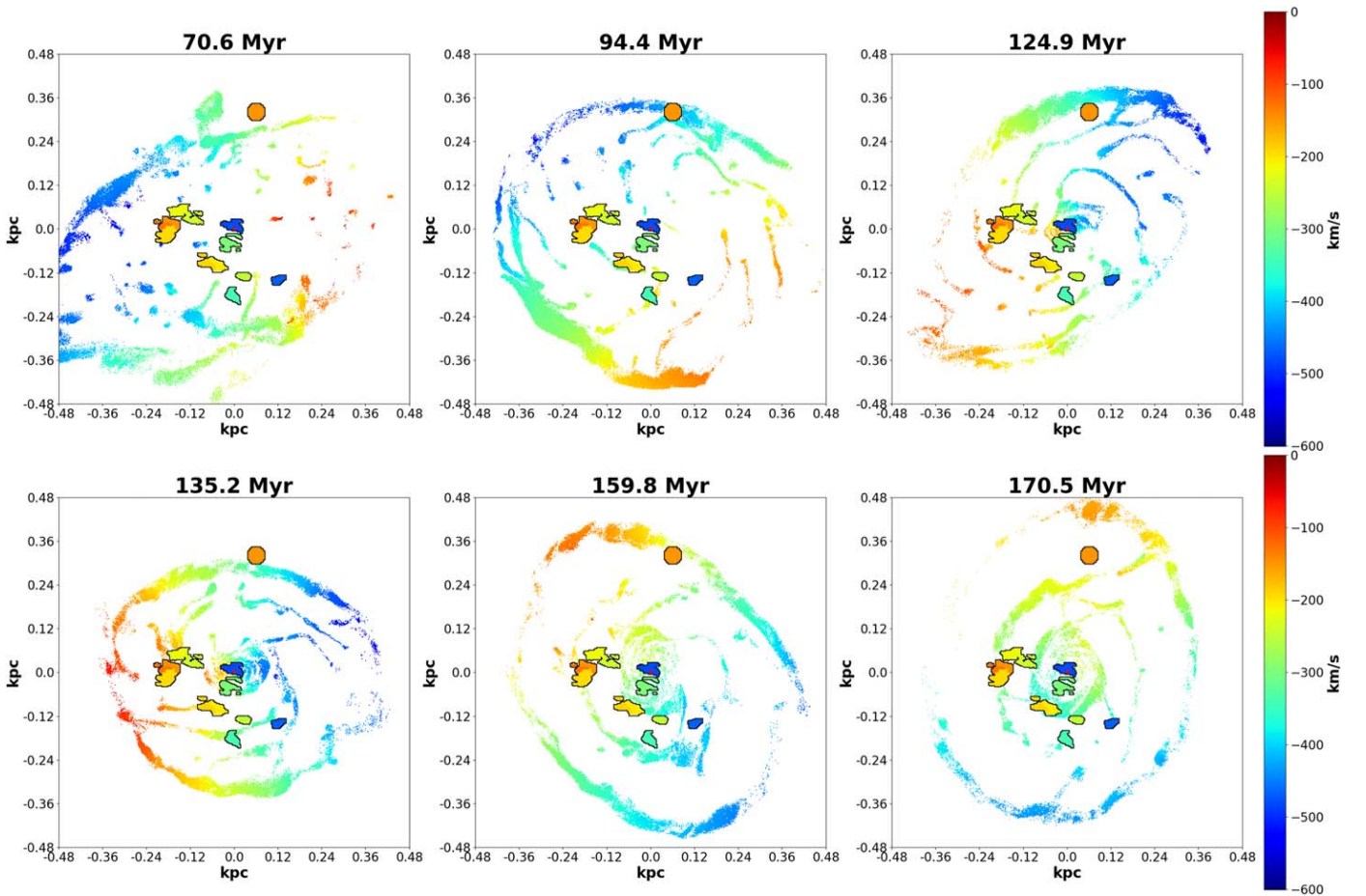

**Figure 10.** Comparison of line-of-sight velocity (in km s$^{-1}$) between observed CO velocities (black outlined patches, color-coded) and gas below $10^4$ K from our simulation. The plane of the resulting disk precesses. We find a reasonable fit (lower middle panel) to the observed velocity field and filament morphology at about 160 Myr of evolution from the initial injection of the streamer into the potential.

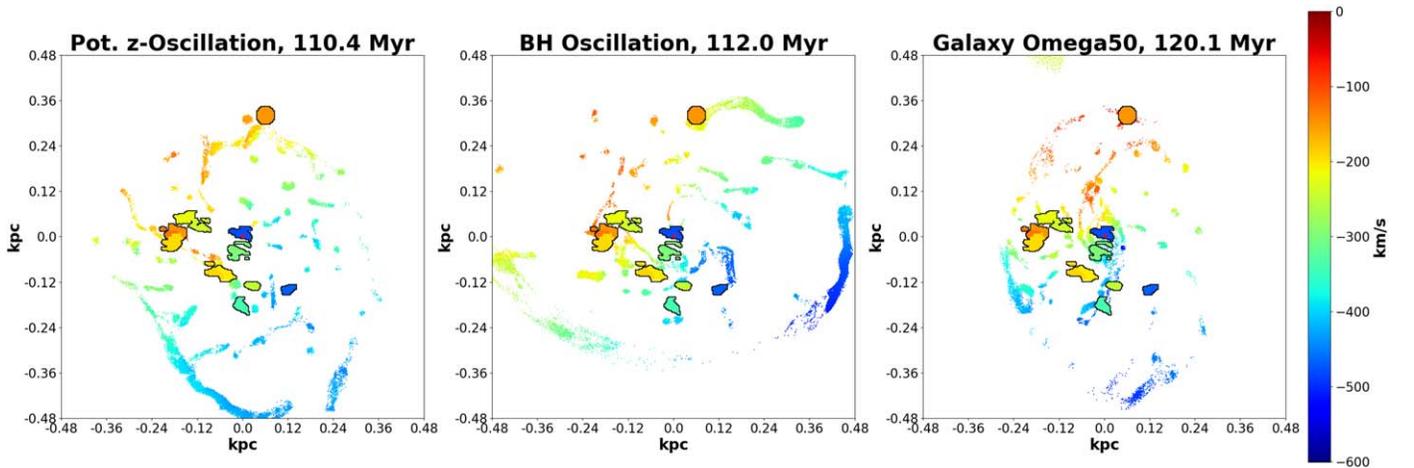

**Figure 11.** Comparison of the line-of-sight velocity (in km s$^{-1}$) between the observed CO data (black patches) and the gas below $10^4$ K from our simulations. The first panel shows a simulation snapshot in which the effect of an external perturbation is approximated by oscillating the M31 potential along the z-axis (line-of-sight direction). Another type of perturbation, shown in the second panel, is approximated by oscillating the BH along the y-axis. In the last panel, the M31 potential rotates at a pattern speed of 50 km s$^{-1}$ kpc$^{-1}$.

To test the possibility and origin of a tilt in the inner region of M31, we have performed a series of special simulations that include additional influences. First, we explore a basic model for an external perturbation (such as the M32 dwarf) by oscillating the non-point part of the M31 potential along an axis with different periods and amplitudes. In another type of simulation, we oscillate only the pointlike (BH and nuclear ring) part of the potential along an axis, also with different parameters for the oscillation.

Finally, we take into account the fact that the evolution time of 200 Myr set in our simulations is comparable to the pattern speed of the M31 galaxy found by Blaña Díaz et al. (2018). According to these authors, models with pattern speeds Ω equal to 30, 40, and 50 km s$^{-1}$ kpc$^{-1}$ (with 40 providing the best fit) correspond to orbital times of 203, 154, and 123 Myr, respectively, which translates to one to about two full orbits in our simulation time. Thus, we also ran simulations in which the potential is no longer static in time.

In the case of the oscillation of the stellar potential, we achieved the best results compared to observations when the oscillation is in the direction of the line of sight (z-axis): first panel in Figure 11. In this example, the amplitude of the oscillation was set to 200 pc in the z-direction with a period of 100 Myr. In this case, the inner plane of about 100 pc is indeed tilted with respect to the outer region, making it possible to obtain a reasonable fit to all the observed velocities except for the blue velocity blob measured on top of the BH at the coordinate origin.

The result in the case of the oscillating BH simulation, shown in the middle panel of Figure 11, uses an amplitude of 100 pc and a period of 50 Myr. In this case, the influence of the BH does not extend far enough to significantly affect the global evolution of the system. Furthermore, an inner disk similar to the one in Figure 4(f) is not formed because the BH moves through the region, accreting the accumulating gas. Of course, this process is numerically amplified by the 6 pc sink radius around the BH set in our simulations. Nevertheless, we would expect a significant effect on an inner disk from the gas accumulating around the BH if it were moving directly through the disk. Thus, this model contradicts the observations, which show the existence of an inner disk (Figure 5), and thus a BH oscillation, if any, should not be large. This, combined with the short range of the BH influence on the general evolution, makes this scenario unlikely.

The last panel of Figure 11 shows the best fit from the simulations using a pattern speed Ω equal to 50. In general, the addition of the rotation of the potential introduces a lot of chaotic motion, with parts of the stream constantly tilting at different radii. In particular, individual cold clumps can break away from the global flow and enter independent orbits. An example is the dark blue blob next to the BH at about 0.06 kpc on the x-axis and −0.04 kpc on the y-axis. Such complex velocity patterns could, however, explain the double peaks in the CO lines seen at some locations in the spectra of Melchior & Combes (2011).

## 4. Summary of the Simulations and Discussion

We present the results of hydrodynamical simulations to explain the origin and fate of the filamentary dust/gas structure in the central 1 kpc of the M31 galaxy. The model assumes that a gas stream originating from a nuclear ring with an outer radius of 1 kpc falls into the central region under the influence of the M31 stellar plus BH potential. The evolution of the streamer is followed over 200 Myr, during which it experiences friction with the surrounding medium, heating, dissipation, and self-collision. The configuration closest to the observations was found at about 160 Myr, when the gas particles settle into a disk of about 100 pc, and numerous filaments emanating from an outer ring of 0.5 kpc radius begin to feed the disk, defining a nuclear spiral configuration (Figure 4(f)).

The robustness of the results was checked by running a series of simulations with different initial conditions. These include different input mass rates and dynamics and the inclusion of external agents to modify the central M31 potential. Interestingly, the main features of the observations remain captured in all cases (Figures 5 and 10). We think that this is a consequence of the relatively symmetric potential effectively operating in the central kiloparsec of M31. In particular, the inclusion of a large perturbation to the potential, such as an external collision of M31 with its companion M32, or the effect of the global pattern speed of M31, gave results similar to those for a static potential—Figure 4(f). A convergent result from all simulations is that the present configuration of M31 nuclear gas and dust represents a short period in a rapidly changing system.

Comparing Figure 4(f) with Figure 5 we find that the outer 0.5 kpc ring in the simulations may indeed represent the observed one, although it has a larger size, 0.7 kpc to 1.0 kpc radius. The ring forms much closer to the center in our simulation, possibly due to the distance at which we start the injection. We think that starting the injection at distances even further than 1 kpc radius could naturally increase the radius of the ring and at the same time lead to the formation of the inner disk in a single event.

On physical grounds, we assume that the simulation case in which the potential is not static but perturbed by the M31 pattern speed is the most physically motivated approach. This is because the timescale for the filaments to reach their final configuration is comparable to the orbital time of M31. The simulation results, in the right panel of Figure 11, indeed show a good agreement with the observed morphology and velocity field.

More generally, as long as the potential is sufficiently symmetric and allows the orbit of a gaseous stream to intersect itself, we expect the formation of a bubble of hot gas. For example, the center of the Milky Way contains large amounts of hot gas (Morris & Serabyn 1996). Mechanisms to explain the origin of this gas include supernovae (Yamauchi et al. 1990) and stellar winds (Chevalier 1992; Quataert 2004). As shown in our simulations, the simple intersection of a gas stream with itself as it enters the core region around the BH provides an additional source of hot gas. Due to the small amount of mass injected in the simulations, of which at most 30% is above $10^6$ K (~1200 $M_\odot$ distributed within the inner 1 kpc), we do not expect any measurable X-ray signal above the background. However, the occurrence of just 100 similar events at earlier times would be sufficient to produce the entire $10^5$ $M_\odot$ of hot gas observed in the central kiloparsec of M31, provided that the formation time is short compared to the cooling time.

Finally, the M31 BH is in the process of receiving some gas infall. The current accreted mass, 200 Myr after the injection of the first material in the central kiloparsec, is about 33 $M_\odot$ in the inner 6 pc. This implies an infall rate of ~$1.7 \times 10^{-7}$ $M_\odot$ yr$^{-1}$, in line with that inferred from $L/L_\mathrm{Edd}$ of M31 (Section 2.1) and consistent with the quiescent state of the M31 BH.

## 5. Conclusion: M31 Simulations in the General Context and the Accretion Mode of sub-Eddington BHs

Parsec-scale surveys of some of the closest low accreting supermassive BHs, i.e., Eddington luminosity ratios $\ll 10^{-3}$—often associated with low-luminosity active galactic nuclei (AGNs)—reveal a common pattern in the dust morphology, formed by narrow, long dust filaments ending in a spiral in the central few hundred parsecs. Detailed cases of these spiral structures show that they are separate nuclear entities extending at most a few hundred parsecs from the center: ~400 pc in NGC 1097 (Prieto et al. 2005, 2021), 200 pc radius in NGC 6951 (Storchi-Bergmann et al. 2007, their Figure 2), ESO 428-G14 (Prieto et al. 2014, their Figure 8; May et al. 2018), and NGC 1566 (Prieto et al. 2021, their Figure 4). The inferred inflow rates at distances of tens of parsecs from the BH are $\lesssim 0.4$ $M_\odot$ yr$^{-1}$ for the above cases. In some cases of even lower Eddington luminosity ratio, e.g.,

M87 and M31, the nuclear spiral is seen in H$\alpha$ gas, extending up to ~250 pc radius in M87 (Ford et al. 1994, their Figure 2(a)), about 400 pc radius in M31 (Figure 3, this work), and in both cases an inner disk of tens of parsecs in size is formed (Walsh et al. 2013 and references therein for M87; this work, Section 1 for M31). The inferred inflow rate is even smaller in these cases, from $10^{-5}$ $M_\odot$ yr$^{-1}$ in M87 (Prieto et al. 2019) to $10^{-7}$ $M_\odot$ yr$^{-1}$ in M31 (this paper). Admittedly, other parsec-scale studies of low accreting BHs do not show nuclear spirals, but this may be due to an unfavorable line of sight when the nuclear spiral plane is close to edge-on, e.g., the linear filamentary dust structure crossing the central 200 pc of the Sombrero galaxy (Prieto et al. 2014).

Examples of similar nuclear dust spirals can be seen in the WFPC2/HST optical–infrared dust maps of nearby Seyfert galaxies by Pogge et al. (2003). As the authors report, the vast majority of the low-luminosity AGNs in their sample also have nuclear spirals extending over several hundred parsecs. Their core luminosity in Eddington luminosity units is not known for most, but the fact that they are low-luminosity AGNs suggests that they may be sub-Eddington cases. Specific LINER cases are discussed in detail in Elmegreen et al. (1998, 2002). These authors suggest that the dust spirals are a consequence of acoustic turbulence in the inner gas disk of these galaxies, and as such could contribute to gas accretion toward the nucleus (Montenegro et al. 1999). Dust maps of galaxies with higher accreting BHs also show a network of dust filaments and lanes at the core, but their global network morphology is much less organized, often consisting of a single long filament that runs straight across the nucleus (e.g., Martini et al. 2003; Prieto et al. 2014; Mezcua et al. 2016).

M31 is another example where a dense network of ionized gas and dust filaments spirals in the central 500 pc around the BH. The hydrodynamical simulations show that the role of these filaments is to transport matter to the center; however, the net amount that they transport to the center is small—a consequence of their extensive interaction with themselves, their surrounding atmosphere, and the ISM over a timescale of several million years. Of course, the net accreted mass depends on the input mass injected into the original streamer; the fact that it is small does not. On the other hand, the input mass in the streamer is constrained by the observed nature of the filaments. Our extensive parameter study with test simulations showed that input masses in the region of a few $10^3$ $M_\odot$ provide an optimal compromise: lower masses led to no formation of diffuse structures and filaments; higher masses triggered star formation in the filaments, which is not observed. None of the nuclear spirals cited above show star formation in their filaments.

Taken together, a more general picture may emerge that goes beyond the case of M31. Hydrodynamical simulations in M31 suggest that nuclear spirals are rather inefficient carriers of mass to the center, and that tight constraints on the input mass are needed to develop the filaments. In the general context, nuclear spirals appear to be ubiquitous around supermassive BHs in the very low sub-Eddington regime, i.e., $L_\mathrm{bol}/L_\mathrm{Edd} \ll 10^{-3}$. We postulate that when dust/gas filaments in the central hundred parsecs of galaxies get to settle in a nuclear spiral configuration, a low accretion mode of the central BH will result.

## Acknowledgments

This research is supported by the Excellence Cluster ORIGINS, which is funded by the Deutsche Forschungsgemeinschaft (DFG, German Research Foundation) under Germany's Excellence Strategy—EXC-2094-390783311. The surface density plots have been created using the publicly available SPH visualization tool SPLASH by D.J. Price (Price 2007). Partial support by Spain I+D+i PROID202101010130 and PID2020-114092GB-I00 (AS and AP). We would like to thank the reviewer for useful comments and constructive criticism.

## ORCID iDs

A. Prieto https://orcid.org/0000-0002-3585-2639
M. Blaña https://orcid.org/0000-0003-2139-0944
A. Burkert https://orcid.org/0000-0001-6879-9822
O. Zier https://orcid.org/0000-0003-1811-8915